\documentstyle[sprocl,twoside]{article}

\def\be{\begin{equation}}
\def\ee{\end{equation}}
\def\bea{\begin{eqnarray}}
\def\eea{\end{eqnarray}}

\newcommand{\p}{\partial}

\def\IB{\relax\hbox{$\inbar\kern-.3em{\rm B}$}}
\def\IC{\relax\hbox{$\inbar\kern-.3em{\rm C}$}}
\def\ID{\relax\hbox{$\inbar\kern-.3em{\rm D}$}}
\def\IE{\relax\hbox{$\inbar\kern-.3em{\rm E}$}}
\def\IF{\relax\hbox{$\inbar\kern-.3em{\rm F}$}}
\def\IG{\relax\hbox{$\inbar\kern-.3em{\rm G}$}}
\def\IGa{\relax\hbox{${\rm I}\kern-.18em\Gamma$}}
\def\IH{\relax{\rm I\kern-.18em H}}
\def\IK{\relax{\rm I\kern-.18em K}}
\def\IL{\relax{\rm I\kern-.18em L}}
\def\IP{\relax{\rm I\kern-.18em P}}
\def\IR{\relax{\rm I\kern-.18em R}}
\def\IZ{\relax{\rm Z\kern-.5em Z}}

\def\p{\partial}
\def\f{\frac}

\def\citebk#1{\hspace{0.9mm}\raisebox{-1.85mm}[0mm][0mm]
  {\Large\cite{#1}}\hspace{-0.1mm}}

\begin{document}
\begin{flushright}
hep-th/0203207 \\
OUTP-02-13-P\\
\end{flushright}

\vspace{0.3 in}
\sloppy
\title{  STRESS ENERGY TENSOR IN ${\mathbf C\!=\!0}$ 
 LOGARITHMIC\\[1mm] CONFORMAL FIELD THEORY}

\author{  IAN I. KOGAN$^{\,a,b,c}$
 ~and ALEXANDER NICHOLS$^{\,a}$
}

\address{
$^a$Theoretical Physics, Department of Physics, Oxford University\\
1 Keble Road, Oxford, OX1 3NP, UK  \\
$^b$IHES, 35 route de Chartres,
91440, Bures-sur-Yvette,  France \\
$^c$Laboratoire de Physique Th\'eorique,
Universit\'e de Paris XI, \\
91405 Orsay C\'edex, France
}
\maketitle
\abstracts{
We discuss the partners of the stress energy tensor and 
their structure in Logarithmic conformal field theories. 
In particular we draw attention to the fundamental differences between 
theories with zero and non-zero central charge. We analyze the OPE
 for $T$, $\bar{T}$ and the logarithmic partners $t$ and $\bar{t}$
 for $c=0$ theories.
}

\vspace{0.3cm}

\tableofcontents

\newpage
{\narrower\sl\noindent This paper is dedicated to the 
memory of \textbf{Michael Marinov}.
One of us (IIK) had a privilege to  know and admire
 Misha Marinov. He was a man  of principles -- in all
aspects of life, not only in theoretical physics.
 The  decision to choose a subject about  Logarithmic Conformal
Field Theories was not  accidental. These theories are 
at the boundary between theories with   unitary
and non-unitary evolution. Perhaps it is not a well known fact  that 
 Misha  wrote pioneering papers in which he used  non-unitary evolution  
 to describe how a  pure state  can  evolve into a mixed    
one\,\cite{Marinov:qs,Marinov:rd}   long time 
before this topic became popular through Hawking's famous 
paper.\cite{Hawking:1982dj} In  these   early papers he discussed the
  phenomenology of $K^{0}\!-\!\bar{K}^{0}$ oscillations in the presence of  
 quantum mixing. Unfortunately his very important papers were not widely 
known (see  however Ref.\ 4  and M.B.~Mensky paper\,\,\cite{mensky} 
in this Volume) and  did not receive the 
recognition they rightfully  deserved.  We hope 
this  subject would have been to his liking. 
}

\section{Introduction}\label{sec:intro}
The study of conformal invariance in two dimensions has been an
extremely interesting and fruitful area of research for the last
twenty years.\cite{Belavin:1984vu}

During the last ten years an interesting class of conformal field
theories (CFTs) has emerged called logarithmic conformal field
theories (LCFTs). In Ref.\,\protect\citebk{Gurarie:1993xq} the concept
of LCFT was introduced and the presence of logarithmic structure in
the operator product expansion was explained by the indecomposable
representations that can occur in the fusion of primary operators.
These occur when there are fields with degenerate scaling dimensions
having a Jordan block structure. It was shown that in any LCFT one of
these degenerate fields becomes a zero norm state coupled to a
logarithmic partner.\cite{Caux:1996nm} This together with another
property -- extra (hidden) symmetries,\cite{Caux:1996nm,Kogan:1996df}
coming from extra conserved currents in our theory, will be important
for our analysis of the stress-energy tensor structure in LCFT.
 The above-mentioned hidden symmetry  means that  there are extra fields
with integer conformal dimensions.  One can even get extra
states with zero dimension which means that we have a theory with a
non-trivial vacuum. These operators play a prominent role in the
Quantum-Hall effect,\cite{Bhaseen:1999nm,Kogan:1999hz} In this case
the descendents of this extra zero dimension operator may form
logarithmic pairs with currents or higher dimension fields. It is
therefore interesting to see what will happen in the case of the
stress tensor itself -- can it have logarithmic partners or not? -- and
will these partners be primary fields or descendents.

In  our previous paper\,\cite{Kogan:2001ku}  we  addressed  this   issue and 
 suggested some kind of classification for LCFT based on the structure
 of the vacuum and the character of the degeneracy of the stress-energy 
tensor.
In particular the structure of the partners to $T$ in LCFTs with
non-zero central charge and LCFTs with zero central charge behave very
differently. The second class, $c=0$ theories, are a very special
sub-class of LCFTs. They are of utmost importance for both disordered
systems and critical strings.  We showed that at $c=0$ in order to
get a non-trivial theory there must exist a state which is orthogonal
to $T$ and is \emph{not} a descendant of any other field. There could
of course also be other states which are descendants but these are not
required by general arguments. The appearance of such a state is
characterised by at least two coefficients when we restricted ourselves
 to holomorphic sector only.
 In Ref.\,\protect\citebk{Kogan:2001ku} we  discussed the
arguments presented in Refs.\,\protect\citebk{Cardy}\,-\protect\citebk{Moghimi-Araghi:2001qn}
concerning the existence of a logarithmic partner $t$ for the
stress-energy tensor. In particular the emergence of logarithmic
behaviour is not universal if we can decompose the theory into a sum
of non-interacting sectors. It is the mixing between these sectors
which makes the theory logarithmic. This issue was previously
discussed in string theory with a ghost-matter mixing 
term.\cite{SUSY30,Kogan:2000nw}

In Ref.\,\protect\citebk{Kogan:2001ku} only the holomorphic sector was
considered. Here we shall include the antiholomorphic sector for
$c=0$ and demonstrate how these become non-trivially mixed.
\section{General properties of  logarithmic operators}

In LCFT there are logarithmic terms in some correlation functions but
the theories are nonetheless compatible with conformal invariance.  An
LCFT appears when several operators, or their descendents
\cite{Kogan:1998fd}, become degenerate. Here we shall discuss the
simplest situation in which only two operators become degenerate and
form a logarithmic pair, denoted by $C$ and $D$.  The OPE of the
stress-energy tensor T with the logarithmic operators $C$ and $D$ is
non-trivial and involves mixing\,\cite{Gurarie:1993xq}
\bea
T(z) C(w,\bar{w}) &\sim&  \frac{h}{(z-w)^2}C(w,\bar{w}) +
\frac{1}{(z-w)} \partial_z C + \dots \,,\label{stress} \\
T(z)D(w,\bar{w}) &\sim& \frac{h}{(z-w)^2} D(w,\bar{w}) + \frac{1}{(z-w)^2}C(w,\bar{w}) + 
\frac{1}{(z-w)} \partial_z  D +
\dots \,,
\nonumber
\eea
where $h $ is the conformal dimension of the
operators with respect to the holomorphic stress-energy tensor $T(z)$.
The OPE with $\bar{T}$ has the same form  but with $\bar{h}$ instead of
$h$; as usual the scaling dimension is $h+\bar{h}$ and the
 spin of the field is $h-\bar{h}$.

It is a consequence of Eq.\,(\ref{stress})
that under a conformal transformation $ z \rightarrow  w= z +
\epsilon(z)$ the  logarithmic pair is  transformed  as
\bea
\delta C &=&\partial_z\epsilon(z)h C + \epsilon(z) \partial_z C + \dots\,,
\nonumber \\
\delta D &=&\partial_z\epsilon(z)(h D +  C) + \epsilon(z)
\partial_z D + \dots\,,
\eea
which  can be written globally as 
\bea
\left(\begin{array}{c}
C(z,\bar{z}) \\ D(z,\bar{z})
\end{array}\right)
= \left(\frac{\partial w}{\partial z}\right)^{\left(\begin{array}{cc}
h & 0 \\
 1 & h
\end{array}\right) }
\left(\frac{\partial \bar{w}}{\partial \bar{z}}\right)^{\left(\begin{array}{cc}
\bar{h} & 0 \\
 1 & \bar{h}
\end{array}\right) }
\left(\begin{array}{c}
C(w,\bar{w}) \\ D(w,\bar{w})
\end{array}\right)\,.
\eea
One can see that even  holomorphic (antiholomorphic) fields with
dimensions $(h,0)$ or $(0,\bar{h})$ are  transformed under 
the action of {\it both}  $T$ and $\bar{T}$, i.e. there is some sort of 
 holomorphic anomaly for logarithmic pair. 

From this conformal transformation one can derive 
the full  two point functions  for the logarithmic 
pair\,\cite{Gurarie:1993xq,Caux:1996nm} 
\begin{eqnarray}
\langle C(x,\bar{x}) D(y,\bar{y})\rangle &=& 
\langle C(y,\bar{y}) D(x,\bar{x}) \rangle  = \frac{\alpha_D}{(x-y)^{2h}
 (\bar{x}-\bar{y})^{2\bar{h}}}\,,\nonumber \\
\langle D(x,\bar{x}) D(y,\bar{y})\rangle &=& 
 \frac{1}{(x-y)^{2h} (\bar{x}-\bar{y})^{2\bar{h}}} 
\left(-2\alpha_D\ln|x-y| + \alpha_D'\right),
\nonumber \\ 
\langle C(x,\bar{x}) C(y,\bar{y})\rangle  &=& 0\,,
\label{CC}
\end{eqnarray}
 where the constant $\alpha_D$ is determined by the normalization of the $D$
operator and the constant $\alpha_D'$ can be changed by the redefinition
 $D \rightarrow  D + C $. 
Note that Eq.\,(\ref{CC}) is absolutely universal and valid in any
number of dimensions, because only the global properties of
conformal symmetry were used to derive it. One can easily generalize
these formulas to the case when there are $n$ degenerate
fields and the Jordan cell is given by an $n \times n$ matrix, in which
case the maximal power of the logarithm will be $\ln^{n-1} z\,$.

\section{Towards the classification of LCFT}
LCFTs can be naturally divided into classes based on the dimension of
the Jordan blocks involved. Here we shall concentrate on the case of
rank 2 (one logarithmic partner) however it is obvious that our results
will generalise to higher rank Jordan cells. It is perhaps still an
interesting problem to understand if there can be a more complicated
structure at higher rank. 

These theories can be further grouped, as we shall show, 
into four distinct categories in which the stress tensor and its partners have different structures:

\begin{itemize}
\item{$c=0$ Theories}
        \begin{itemize}
        \item 0A:~Non-degenerate vacua ($SU(2)_0$, Disordered Models)
        \item 0B:~Degenerate vacua ($OSp(2|2)_k$ for certain $k$)
        \end{itemize}
\item{$c \ne 0$ Theories}
        \begin{itemize}
        \item IA:~Non-degenerate vacua 
        \item IB:~Degenerate vacua ($c_{p,1}$).
        \end{itemize}
\end{itemize}
Throughout this paper we use the notation that non-degenerate and
degenerate refer to the single vacuum and the vacuum with logarithmic
pair respectively. There may also be other primaries at $h=0$ with a
trivial Jordan cell structure and we do not consider this possibility
here.

Only in the case of the $c_{p,1}$ models \cite{Flohr:1996ea} and in particular the $c=-2$
triplet model has the structure of the theory been fully elucidated \cite{Gaberdiel:1999ps}.
For the others some of the structure is known from explicit
correlation functions. As far as we are aware there has been no
examples of type IA in the literature.  It is easy to see that a
logarithmic partner for $T$ can only exist in cases $0A$, $0B$, $IB$ by the
following simple arguments.

If $T$ has a logarithmic partner then $T$ itself must be a zero norm 
state\,\cite{Caux:1996nm}
\bea \label{eqn:Tzeronorm}
\left< T(z) T(w) \right> =0\,.
\eea
Now consider the standard OPE for the stress tensor
\bea
T(z) T(w) \sim \f{c ~I}{2(z-w)^4}+\f{2 T(w)}{(z-w)^2}+\f{\p T(w)}{z-w} + \cdots
\eea
where $I$ is the identity operator. For consistency with (\ref{eqn:Tzeronorm}) we see that we must have
%
\bea
c \left< I \right> =c \left< 0|I|0 \right> = c\left< 0|0 \right>=0\,.
\eea
For $c \ne 0$ this implies that the vacuum $\left.|0\right>$ must have zero norm and thus must be part of a logarithmic pair which excludes case IA. Thus partners to the stress tensor $T$ cannot occur in type IA theories. For this reason in the next section when discussing non-degenerate vacua we shall only discuss the case of $c=0$.

\section{Non-degenerate vacua and ${\mathbf c \rightarrow 0}$ limit}

\subsection{$c=0$ catastrophe}

Here we review the construction given in 
Refs.\,\protect\citebk{Cardy}, \protect\citebk{Gurarie:1999yx}, 
\protect\citebk{CardyTalk}.

For a primary field of conformal dimension $h$ we use the normalisation
\bea
\left< V(z_1,\bar{z}_1) V(z_2,\bar{z}_2) \right> = \f{A}{z_{12}^{2h}\bar{z}_{12}^{2 \bar{h}}}\;.
\eea
Then  we consider the correlator
\bea
\left< T(z) V(z_1,\bar{z}_1) V(z_2,\bar{z}_2) \right> = \f{A~h}{(z-z_1)^2(z-z_2)^2 z_{12}^{2h-2}\bar{z}_{12}^{2 \bar{h}}  }\;.
\eea
The coefficient of the three point function is uniquely fixed by considering the limit $z \rightarrow z_1$ and using the property of a primary field,
\bea
T(z) V(w,\bar{w}) \sim \f{h V(w,\bar{w})}{(z-w)^2}+\f{\p
V(w,\bar{w})}{z-w} +
 \cdots\;.
\eea
Of course we have similar results following from insertions of $\bar{T}(\bar{z})$ in the correlator and we have taken $h=\bar{h}$ for simplicity. We now use
\bea
\left< T(z) T(0) \right> = \f{c}{2 z^4}\,, \quad
 \left<0|I|0 \right>= \f{c}{2 z^4}\,; \qquad  \left<0|0 \right> =1\,, 
\eea
and are explicitly using the fact that the identity field has non-zero norm. If $T$ is the only $h=2$ field present in our model we can deduce
\bea
V(z,\bar{z}) V(0,0) \sim \f{A(c)}{z^{2h}\bar{z}^{2 h}} \left[ 1 +  \f{2h}{c} z^2 T(0) + \f{2 \bar{h}}{c} \bar{z}^2 \bar{T}(0) +  \cdots \right].
\eea
We have also assumed $c=\bar{c}$.
Clearly for $c=0$ if $A(0) \ne 0$ then the above OPE becomes ill-defined.
However suppose that as $c$ approaches zero there is another spin $0$ field $X$ with dimension $(2+\alpha,\alpha)$ and approaches $(2,0)$. For most of the following we shall concentrate on the operators with dimensions that converge to $(2,0)$ however a similar pattern occurs for the $(0,2)$ operators. Then for $c \ne 0$ we have
\bea \label{eqn:OPE}
V(z,\bar{z}) V(0,0) \sim \f{A(c)}{z^{2h}} \left[ 1 +  \f{2h}{c} z^2 T(0)
+ 2 X(0,0) z^{2+\alpha(c)} \bar{z}^{\alpha(c)} + \cdots \right].
\eea
%


Our starting point will be the two point functions for $c \ne 0\,$. Non-chiral fields $X(z,\bar{z})$, $\bar{X}(z,\bar{z})$ are of dimensions $(2+\alpha,\alpha)$ and $(\alpha,2+\alpha)\,$, respectively. The only non-trivial 2-point correlators (up to relation by conjugation) are
\bea \label{eqn:XX}
\left< T(z_1) T(z_2) \right> &=& \f{c}{z_{12}^4}\;, \\
\left< X(z_1,\bar{z}_1) X(z_2,\bar{z}_2) \right>&=& \f{1}{c}\,\f{B(c)}{z_{12}^{4+2\alpha(c)}\bar{z}_{12}^{2 \alpha(c)}}\;,
\eea
where we used the fact that $\left<T(z_1) X(z_2,\bar{z}_2) \right>$ vanishes as they have different dimensions.
We have exactly the same relations for the fields related by conjugation.
We define the new fields $t$, $\bar{t}$ via
\bea
t= \f{b}{c}T+\f{b}{h} X, ~ 
 \hskip 1 cm \bar{t}= \f{b}{c}\,\bar{T}+\f{b}{h}\, \bar{X}\,.
\eea
The parameter $b$ is defined through
\bea \label{eqn:bdef}
b^{-1} \equiv -\lim_{c \rightarrow 0} \f{\alpha(c)}{c}=-\alpha'(0).
\eea
We can now calculate the two point function,
\bea
\left< T(z_1) t(z_2,\bar{z}_2) \right> &=& 
\left< T(z_1) \left[ \f{b}{c} T + \f{b}{h}  X \right](z_2,\bar{z}_2)
\right>  
 = \f{b}{c} \left< T(z_1) T(z_2) \right> \nonumber \\
&=& \f{b}{2}\f{1}{ z_{12}^4}.
\eea
Also,
\bea
\left< t(z_1,\bar{z}_1) t(z_2,\bar{z}_2) \right> &=& 
 \left< \left[ \f{b}{c} T + \f{b}{h}  X \right](z_1,\bar{z}_1) \left[ \f{b}{c} T + \f{b}{h}  X \right](z_2,\bar{z}_2) \right> \nonumber \\
&=& \f{b^2}{c^2} \left< T(z_1) T(z_2) \right> +
 \f{b^2}{h^2} \left< X(z_1,\bar{z}_1) X(z_2,\bar{z}_2) \right> \nonumber \\
&=&\f{b^2}{2c}\f{1}{z_{12}^4}+\f{b^2 B(c)}{h^2 c} \f{1}{z_{12}^{4+2\alpha(c)} \bar{z}_{12}^{2 \alpha(c)}}  \\
&=&\f{b^2}{2c}\f{1}{z_{12}^4}+\f{b^2 B(c)}{h^2 c} \f{1}{z_{12}^4} \left( 1- 2\alpha(c) \ln |z_{12}|^2 + \cdots \right)\nonumber \\
&=&\f{1}{z_{12}^4} \left\{ \left( \f{b^2}{2c}+\f{b^2 B(c)}{h^2 c} \right) - \f{2b^2 B(c)\alpha(c)}{h^2c}\ln |z_{12}|^2 + \cdots \right\}. \nonumber
\eea
As this is to be well defined we see that we must have
$B(c)\!=\!-\f{1}{2} h^2\!+\!B_1c\! +\!O(c^2) $. Now using (\ref{eqn:bdef}) 
we get the standard OPEs for a logarithmic pair,
\bea
\left< T(z_1) T(z_2) \right>&=&0\;, \nonumber \\
\left< T(z_1) t(z_2,\bar{z}_2) \right>&=& \f{b}{2 z_{12}^4}\;, \\
\left< t(z_1,\bar{z}_1) t(z_2,\bar{z}_2) \right>&=& 
\f{B_1 -b\ln |z_{12}|^2}{z_{12}^4}\;. \nonumber
\eea
As discussed earlier $B_1$ can be removed by a redefinition of $t$ and
we shall assume that this has been done. Note that although $t$ is a
$(2,0)$ field it is not chiral as $\bar{\p} t \ne 0$. We have similar
expressions for the correlation functions related to these by
conjugation. Also,
\bea
\left< T(z_1) \bar{t}(z_2,\bar{z}_2) \right>\!\!\!&=&\!\!\!
\left< T(z_1) \left[ \f{b}{c} +\f{b}{h} X \right] (z_2,\bar{z}_2)
\right>=0\;,
 \\
\left<  t(z_1,\bar{z}_1) \bar{t}(z_2,\bar{z}_2)\!\!\! \right>&=&\!\!\!
\left<  \left[ \f{b}{c} +\f{b}{h} X \right] (z_1,\bar{z}_1) 
\left[ \f{b}{c} +\f{b}{h} X \right] (z_2,\bar{z}_2) \right>=0\;.
\eea

The OPE (\ref{eqn:OPE}) now becomes,
\bea 
V(z,\bar{z}) V(0,0) \sim \f{A(0)}{z^{2h}} \left[ 1\! + \!
 \f{2h}{b} z^2 \left( T  \ln |z|\!+\! t \right)\! + \! \f{2h}{b} \bar{z}^2 
\left( \bar{T} \ln |z|\!+\! \bar{t} \right)\! +\! \cdots \right],
\label{logOPE}
\eea
which now only involves quantities that are perfectly well defined in the limit as $c \rightarrow 0$.

We can now continue and insist that $t$ is also well defined in the
three point functions (See Appendix). Assuming now that the algebra
closes these are then sufficient to determine the OPEs. These are, 
\bea
&& T(z_1) t(z_2,\bar{z}_2) 
\sim \f{b}{2 z_{12}^4} + \f{2 t(z_2,\bar{z}_2) +
T(z_2)}{z_{12}^2} + \f{\p t(z_2,\bar{z}_2) }{z_{12}}+ \cdots \;,\\[1mm]
%
&& T(z_1) \bar{t}(z_2,\bar{z}_2)  \sim
 \f{\bar{T}(\bar{z}_2)}{z_{12}^2} +
\f{\p \bar{t}(z_2,\bar{z}_2)}{z_{12}}+ \cdots  \;,\\[1mm]
%
&& t(z_1,\bar{z}_1) t(z_2,\bar{z}_2) \sim   \f{-b
\ln|z_{12}|^2}{z_{12}^4}
 + \f{1}{z_{12}^2} 
\Bigg[ \left( 1-4 \ln |z_{12}|^2 \right) t(z_2,\bar{z}_2)\nonumber\\ 
&&  
\hspace*{2.5cm} +\left( \f{2a}{b} - \ln|z_{12}|^2 -2\ln^2 |z_{12}|^2 \right) T(z_2)
\Bigg]   \\
&&\hspace*{2.5cm}+ \,\f{\bar{z}_{12}^2}{z_{12}^4} \left[ \bar{t}(z_2,\bar{z}_2) 
+ 
\left( \f{2f}{b} + \ln|z_{12}|^2 \right) \bar{T}(\bar{z}_2) \right] + 
\cdots\;, \nonumber\\[1mm]
%
&&t(z_1,\bar{z}_1) \bar{t}(z_2,\bar{z}_2) \sim \f{1}{\bar{z}_{12}^2} \left[ \left( \f{2f}{b} - \ln|z_{12}|^2 \right) T(z_2) + \cdots \right] \\[1mm]
&&\hspace*{2.5cm}+\,\f{1}{z_{12}^2} \left[ \left( \f{2f}{b}-\ln|z_{12}|^2 \right) \bar{T}(\bar{z}_2) + \cdots \right].  \nonumber 
\eea
The appearance of a state $\left.|t\right> = t\left.|0\right> $ in this way 
is equivalent to postulating a logarithmic partner for the null vector
$T$.  This prevents $T$ from 
decoupling despite the fact that it is a zero-norm state.
Note that once one fixes
\bea
L_{0}\left.|t\right>&=&2\left.|t\right>+\left.|T\right>\,, \\[1mm]
\bar{L}_0 \left.|t\right>&=&\left.|T\right>
\eea
then the parameter $b$ cannot be removed by rescaling and thus different values of $b$ correspond to inequivalent representations.

Let us note that in our notation parameter $b$ is different by a factor of 2 from a definition given in Ref.\,\protect\citebk{Gurarie:1999yx}. We also want to stress that  the $t(z) t(0) $ OPE 
 is determined by several parameters, not by only $b$ as in Ref.\,\protect\citebk{Gurarie:1999yx}. 
The constant terms  $\f{2a}{b}$, $\f{2f}{b}$ cannot be removed by scale transformation -- one can absorb it in $\ln |z|$ term, but 
 not into $\ln^2\! |z|\,$. 
 One of the important open problems  is to  find  if  the classification of all $c=0$ theories of this type can be 
reduced to the classification of all possible triplets $(a,b,f)$. 
The parameter $a$ cannot be  
 determined from singular terms in  (\ref{logOPE}), but only from the full 3-point
 functions of $(T,t)$ pair (see Appendix).
\subsection{$c=0$ and separability}

There is a third rather trivial way out of the paradox at $c=0$. It is
simply that the full theory is constructed from two parts $T=T_1
\oplus T_2,c=c_1+c_2=0$ both having $c_i \ne 0$. Then in the OPE of
two fields from one part we will only see the stress tensor for that
part rather than the full one. Operators in the full theory are just
the direct product $V=V_1 \otimes V_2$. Then (writing only the holomorphic
 part)
\bea
&&V(z) V(0) = V_1(z) V_1(0) ~~ V_2(z) V_2(0)  \\[1mm]
&&\sim  \f{1}{z^{2h_1}} \left( 1+ z^2 \f{2h_1}{c_1}T_1(0) + \cdots
\right) 
 \f{1}{z^{2h_1}} \left( 1+ z^2 \f{2h_2}{c_2}T_2(0) + \cdots \right)
\nonumber
 \\[1mm]
&&\sim \f{1}{z^{2h}} \left[ 1+ z^2 \left( \f{2h_1 T_1}{c_1} 
+ \f{2h_2 T_2}{c_2} \right) +\cdots \right]. \nonumber 
\eea
This expression is now perfectly well defined as $c_1,c_2 \ne 0$. Of course 
this is as expected as the two decoupled theories are perfectly regular.

In critical string theory the ghost and matter sectors are normally assumed 
to be non-interacting. However this may not be the most general if we wish 
to allow not just positive but 
also zero norm states in our final theory.\cite{SUSY30}

\section{Conclusions}

We   studied here  the stress tensor and its partners in LCFT. 
In particular in non-trivial $c=0$ theories we have demonstrated that it 
is necessary for there to be a primary field of dimension $2$ orthogonal
 to the stress tensor. 
The indecomposable representations are characterised by at least three
 parameters: $a$, $b$ and $f$. The third parameter $f$ emerges when we 
 include antiholomorphic sector.  In a logarithmic theory one can not
avoid mixing between holomorphic and antiholomorphic sectors.

\section*{Acknowledgements}
\addcontentsline{toc}{section}{\numberline{}Acknowledgements}
We would like to thank J.~Cardy and V.~Gurarie for interesting comments.
A.N.\ is funded by the Martin Senior Scholarship, Worcester College,
Oxford. I.I.K.\ is partly supported by PPARC rolling grant
PPA/G/O/1998/00567 and EC TMR grant HPRN-CT-1999-00161.

\section*{Appendix: Three point functions}
\addcontentsline{toc}{section}{\numberline{}Appendix: Three point functions}
We now consider the three point functions. They are the following:
\bea 
&&\!\!\!\!\!\left< T(z_1) T(z_2) T(z_3) \right> = \f{c}{z_{12}^2 z_{13}^2 z_{23}^2 }\;, 
\label{eqn:OPEs}
\\[0.5mm]
&&\!\!\!\!\!\left< T(z_1) X(z_2,\bar{z}_2) X(z_3,\bar{z}_3) \right>
=\f{1}{c}\,\f{C(c)}{z_{12}^2 z_{13}^2 z_{23}^{2+2\alpha(c)}
\bar{z}_{23}^{2\alpha(c)} }\;,\nonumber
\\[0.5mm]
&&\!\!\!\!\!\left< X(z_1,\bar{z}_1) X(z_2,\bar{z}_2) X(z_3,\bar{z}_3) \right> =
\f{1}{c^2}\f{D(c)}{z_{12}^{2+\alpha(c)} z_{13}^{2+\alpha(c)} 
z_{23}^{2+\alpha(c)} \bar{z}_{12}^{\alpha(c)}
 \bar{z}_{13}^{\alpha(c)} \bar{z}_{23}^{\alpha(c)} }\;,\nonumber
\\[0.5mm]
&&\!\!\!\!\!\left< T(z_1) \bar{X}(z_2,\bar{z}_2) \bar{X}(z_3,\bar{z}_3) \right> =
\f{E(c)}{z_{12}^2 z_{13}^2 z_{23}^{2 \alpha(c)-2}
\bar{z}_{23}^{4+2\alpha(c)}}\;,\nonumber
\\[0.5mm]
&&\!\!\!\!\!\left< X(z_1,\bar{z}_1) X(z_2,\bar{z}_2) \bar{X}(z_3,\bar{z}_3) \right>
= 
\f{1}{c}\f{F(c)}{z_{12}^{4+\alpha(c)} z_{13}^{\alpha(c)}
z_{23}^{\alpha(c)} \bar{z}_{12}^{\alpha(c)-2} \bar{z}_{13}^{2+\alpha(c)} 
\bar{z}_{23}^{2+\alpha(c)} }\;.
 \nonumber
\eea
We note that all correlators are single-valued for any $\alpha(c)$ and
therefore must also be at the critical point. This is important as
logarithmic terms should only emerge in the form $\ln|z|$.

We have already fixed the normalisation of the two point functions
(\ref{eqn:XX}). Then by expanding the three point functions we see
that
\bea \label{eqn:known}
C(c)&=&(2+\alpha(c)) B(c)=(2+\alpha(c))\left(-\f{1}{2} h^2+B_2c^2 +\cdots\,\right)\,, \\
E(c)&=&\f{\alpha(c)}{c} B(c)= \f{\alpha(c)}{c}  (-\f{1}{2} h^2+B_2c^2 
+\cdots\,)\;. 
\eea

As we wish to have well defined operators $T,\;t,\;\bar{T},\; \bar{t}\,$ they must in particular have regular 3-point functions. This will be enough to determine the leading behaviour of the functions above.
Consider
\bea
\left< T(z_1) T(z_2) t(z_3,\bar{z}_3) \right> = \left< T(z_1) T(z_2) \left[ \f{b}{c}T+\f{b}{h} X \right] (z_3,\bar{z}_3) \right> 
= \f{b}{z_{12}^2 z_{13}^2 z_{23}^2 }\,.
\eea
%
%
\vspace{-3mm}
\bea
\hspace{-1cm}\left< T(z_1) t(z_2,\bar{z}_2) t(z_3,\bar{z}_3) \right>  
&=& \left< T(z_1) 
 \left[ \f{b}{c}T+\f{b}{h} X \right] (z_2,\bar{z}_2) \left[ \f{b}{c}T+\f{b}{h} X \right] (z_3,\bar{z}_3) \right> \nonumber\\
&=& 
\f{b^2}{c^2}\f{c}{z_{12}^2 z_{13}^2 z_{23}^2 } +
\f{b^2}{h^2 c}\f{C(c)}{z_{12}^2 z_{13}^2 z_{23}^{2+2\alpha(c)}
\bar{z}_{23}^{2\alpha(c)}  }
\\
&=& \f{b^2}{z_{12}^2 z_{13}^2 z_{23}^2 } \left[ \f{1}{c} + \f{C(c)}{h^2 c}\left(1-2\alpha(c) \ln |z_{23}|^2 +\cdots \right) \right]. \nonumber
\eea
Now using the form of $C(c)$ (\ref{eqn:known}) we get
\bea
\left< T(z_1) t(z_2,\bar{z}_2) t(z_3,\bar{z}_3) \right>
 =\f{-2b\ln |z_{23}|^2+\f{b}{2}}{z_{12}^2 z_{13}^2 z_{23}^2 }\;,
\eea
and 
\bea
&&\left< t(z_1,\bar{z}_1) t(z_2,\bar{z}_2) t(z_3,\bar{z}_3) \right>  
\nonumber \\[1mm]
&& =\left<  \left[ \f{b}{c}T+\f{b}{h} X \right] (z_1,\bar{z}_1)   \left[ \f{b}{c}T+\f{b}{h} X \right] (z_2,\bar{z}_2) \left[ \f{b}{c}T+\f{b}{h} X \right] (z_3,\bar{z}_3) \right>  \nonumber  \\[1mm]
&&= \f{b^3}{c^3} \left< T(z_1) T(z_2) T(z_3) \right> + \f{b^3}{h^2 c} \left( \left< X(z_1,\bar{z}_1) X(z_2,\bar{z}_2) T(z_3) \right> \right.  \nonumber  \\[1mm]
&&\left. + \left< X(z_1,\bar{z}_1) T(z_2) X(z_3,\bar{z}_3) \right> + \left< T(z_1) X(z_2,\bar{z}_2) X(z_3,\bar{z}_3) \right> \right) \\[1mm]
&&+\f{b^3}{h^3} \left< X(z_1,\bar{z}_1) X(z_2,\bar{z}_2) X(z_3,\bar{z}_3) \right>  \nonumber  \\[1mm]
&&=\f{b^3}{c^2} \f{1}{z_{12}^2 z_{13}^2 z_{23}^2 }
 + \f{b^3}{h^3 c^2} \f{D(c)}{z_{12}^2 z_{13}^2 z_{23}^2 }
z_{12}^{-\alpha(c)} z_{13}^{-\alpha(c)} z_{23}^{-\alpha(c)} \nonumber   \\[1mm]
&&+ \f{b^3}{h^2 c^2} \f{C(c)}{z_{12}^2 z_{13}^2 z_{23}^2 } \left
[ z_{12}^{-2\alpha(c)}\bar{z}_{12}^{-2\alpha(c)} +
z_{13}^{-2\alpha(c)}\bar{z}_{13}^{-2\alpha(c)} +
z_{23}^{-2\alpha(c)}\bar{z}_{23}^{-2\alpha(c)} \right]. \nonumber 
\eea
Now expanding this and using (\ref{eqn:bdef}):
\bea
&&\left< t(z_1) t(z_2) t(z_3) \right>  = \f{b^3}{h^3 c^2}\left(-2h^3+D_0\right) \\
&&+ \f{b^2}{h^3 c} \left[ \left(D_0-2h^3)(\ln |z_{12}|^2+ \ln |z_{13}|^2+ \ln |z_{23}|^2 \right) + bD_1+\f{3}{2}h^3 \right] +O(1).  \nonumber 
\eea
Thus if this is to be regular in the limit we must have
\bea
D_0=2h^3, ~~~~~ D_1=-\f{3h^3}{2b}.
\eea
Then from the $O(1)$ terms we get
\vspace{-3mm}
\bea
&&\!\!\!\!\!\!\!\!\!\left< t(z_1,\bar{z}_1) t(z_2,\bar{z}_2) t(z_3,\bar{z}_3) \right>  = 
\f{1}{z_{12}^2 z_{13}^2 z_{23}^2 } 
\Biggl\{ -b\left( \ln^2 |z_{12}|^2 +  \ln^2 |z_{13}|^2 +  \ln^2
|z_{23}|^2 \right) \nonumber   \\ 
&&+ 2b \left(\ln |z_{12}|^2 \ln |z_{13}|^2 + \ln |z_{12}|^2 \ln |z_{23}|^2 + \ln |z_{13}|^2 \ln |z_{23}|^2 \right)   \\
&& - \f{b}{2} \left( \ln |z_{12}|^2 +  \ln |z_{13}|^2 +  \ln |z_{23}|^2  \right) + a \Biggr\},  \nonumber
\eea
where we have defined the constant $a$ by
\bea
a \equiv -\f{b^3}{2h^3}\left( -2 D_2 -12hB_2 +\f{3}{2}h^3 \alpha''(0) \right).
\eea
Now consider correlators involving the $\bar{T},\bar{X}$ fields as well.
 For instance
\bea
&&\left< T(z_1) T(z_2) \bar{t}(z_3,\bar{z}_3) \right> = \left< T(z_1) T(z_2) \left[ \f{b}{c}\bar{T} + \f{b}{h}\bar{X} \right] (z_3,\bar{z}_3) \right> =0, \\
&&\left< T(z_1) \bar{T}(\bar{z}_2) t(z_3,\bar{z}_3) \right> = \left< T(z_1) \bar{T}(\bar{z}_2) \left[ \f{b}{c}T + \f{b}{h} X \right] (z_3,\bar{z}_3) \right> =0,  \nonumber\\
&&\!\!\!\!\!\!\!\!\!\left< T(z_1) t(z_2,\bar{z}_2) \bar{t}(z_3,\bar{z}_3) \right>\! =\!
 \left< T(z_1)\! \left[ \f{b}{c} T\! +\! \f{b}{h} X \right] (z_2,\bar{z}_2) \left[ \f{b}{c}\bar{T}\! +\! \f{b}{h} \bar{X} \right]\! (z_3,\bar{z}_3) \right>\! =\!0\,.  \nonumber
\eea
More non-trivially
\bea
&&\!\!\!\!\!\!\!\!\!\left< T(z_1) \bar{t}(z_2,\bar{z}_2) \bar{t}(z_3,\bar{z}_3) \right> =
 \left< T(z_1) \left[ \f{b}{c} \bar{T} + 
\f{b}{h} \bar{X} \right] (z_2,\bar{z}_2) \left[ \f{b}{c}\bar{T} + 
\f{b}{h} \bar{X} \right] (z_3,\bar{z}_3) \right>   \nonumber\\
&&\hspace*{3cm}\;= \f{b^2}{h^2} \,\f{E(c)}{z_{12}^2 z_{13}^2 z_{23}^{2 \alpha(c)-2} \bar{z}_{23}^{4+2\alpha(c)}}\;.
\eea
Inserting the known expression  for of $E(c)$ we get
\bea
\left< T(z_1) \bar{t}(z_2,\bar{z}_2) \bar{t}(z_3,\bar{z}_3) \right> =  
\f{\f{b}{2}}{z_{12}^2 z_{13}^2 z_{23}^{-2} \bar{z}_{23}^{4}}\;.
\eea
The last correlator we have to consider is the following:
\bea
&&\left< t(z_1,\bar{z}_1) t(z_2,\bar{z}_2) \bar{t}(z_3,\bar{z}_3) \right> 
\\[1mm]
&&=\left< \left[ \f{b}{c} T +\f{b}{h} X \right](z_1,\bar{z}_1) \left
[ \f{b}{c} T +\f{b}{h} X \right](z_2,\bar{z}_2) \left[ \f{b}{c} \bar{T}
+\f{b}{h} 
\bar{X} \right](z_3,\bar{z}_3) \right>  \nonumber \\
&&=\f{b^3}{c h^2} \left< X(z_1,\bar{z}_1) X(z_2,\bar{z}_2) 
\bar{T}(z_3,\bar{z}_3) \right> + \f{b^3}{h^3} \left< X(z_1,\bar{z}_1) 
X(z_2,\bar{z}_2) \bar{X}(z_3,\bar{z}_3) \right> \nonumber  \\
&&=\f{b^3 E(c)/c h^2}{z_{12}^{4+2 \alpha(c)} \bar{z}_{12}^{2
\alpha(c)-2} \bar{z}_{13}^2 \bar{z}_{23}^2 } 
+\f{b^3 F(c)/c h^3}{z_{12}^{4+\alpha(c)} z_{13}^{\alpha(c)}
z_{23}^{\alpha(c)} \bar{z}_{12}^{\alpha(c)-2} \bar{z}_{13}^{2+\alpha(c)}
\bar{z}_{23}^{2+\alpha(c)} }\;. \nonumber
\eea
Thus we find
\vspace{-2mm}
\bea
F(c)=-\f{h^3}{2b}+F_1 c + O(c^2).
\eea
Finally we get
\bea
&&\!\!\!\!\!\!\!\!\!\!\!\!\left< t(z_1,\bar{z}_1) t(z_2,\bar{z}_2) \bar{t}(z_3,\bar{z}_3) \right>
=\f{ \f{b}{2} \left( \ln|z_{12}|^2-\ln|z_{13}|^2-\ln|z_{23}|^2 \right) + f}{z_{12}^4 \bar{z}_{12}^{-2} \bar{z}_{13}^2 \bar{z}_{23}^2 } \,,
\eea
where the coefficient $f=-b^3(-\f{1}{2}h^3\alpha''(0) -2F_1)/(2h^3)$. 
%

In summary we have found the following correlators which yield the OPEs given in the text:
\bea
\left< T(z_1) t(z_2,\bar{z}_2) \right>\!\!&=&\!\! \f{b}{2 z_{12}^4} \;,\\
\left< t(z_1,\bar{z}_1) t(z_2,\bar{z}_2) \right> \!\!&=&\!\!
\f{-b\ln |z_{12}|^2}{z_{12}^4} \;,\\
\left< T(z_1) T(z_2) t(z_3,\bar{z}_3) \right>\!\!&=&\!\!
\f{b}{z_{12}^2 z_{13}^2 z_{23}^2 } \;,\\
\left< T(z_1) t(z_2,\bar{z}_2) t(z_3,\bar{z}_3) \right> \!\!&=&\!\!
\f{-2b\ln |z_{23}|^2+\f{b}{2}}{z_{12}^2 z_{13}^2 z_{23}^2 }\;, 
\eea
\vspace{-5mm}
\bea
 && \hspace{-8mm} \left< t(z_1,\bar{z}_1) t(z_2,\bar{z}_2) t(z_3,\bar{z}_3) \right>  =
\f{1}{z_{12}^2 z_{13}^2 z_{23}^2 } \Biggl\{ -b\left( \ln^2 |z_{12}|^2 + 
 \ln^2 |z_{13}|^2 +  \ln^2 |z_{23}|^2 \right)  \nonumber\\
&&\hspace{20mm} + 2b \left(\ln |z_{12}|^2 \ln |z_{13}|^2 +
 \ln |z_{12}|^2 \ln |z_{23}|^2 + \ln |z_{13}|^2 \ln |z_{23}|^2 \right)  \nonumber \\
&& \hspace{20mm}- \f{b}{2} \left( \ln |z_{12}|^2 +  \ln |z_{13}|^2 +  \ln |z_{23}|^2  
\right) + a \Biggr\} , \\
&&\hspace{18mm} \left< T(z_1) \bar{t}(z_2,\bar{z}_2) \bar{t}(z_3,\bar{z}_3) \right> =
 \f{\f{b}{2}}{z_{12}^2 z_{13}^2 z_{23}^{-2} \bar{z}_{23}^{4}} \;,\\
&&\hspace{-3mm}\left< t(z_1,\bar{z}_1) t(z_2,\bar{z}_2) \bar{t}(z_3,\bar{z}_3) \right>
=  \f{\f{b}{2} \left( \ln|z_{12}|^2-\ln|z_{13}|^2-\ln|z_{23}|^2
\right) + 
f }{z_{12}^4 \bar{z}_{12}^{-2} \bar{z}_{13}^2 \bar{z}_{23}^2 } \;.
\eea
%
\\[2mm]
{\bf References}
\addcontentsline{toc}{section}{\numberline{}References}

\end{document}